\documentclass{article}
\usepackage{spconf,amsmath,graphicx}
\usepackage{glossaries}
\usepackage{amssymb}
\usepackage{subcaption}
\usepackage{tikz}
\usetikzlibrary{shapes.arrows}
\usetikzlibrary{shapes, arrows, positioning}
\usetikzlibrary{decorations.pathreplacing}
\usetikzlibrary{plotmarks}
\usepackage[]{fancyhdr}
\usetikzlibrary{patterns}

\newacronym{iot}{IoT}{Internet of Things}
\newacronym{ml}{ML}{Machine Learning}
\newacronym{mi}{MI}{Mutual Information}
\newacronym{ez}{EZ}{Epileptogenic Zone}
\newacronym{soz}{SOZ}{Seizure Onset Zone}
\newacronym{seeg}{SEEG}{Stereo-Electroencephalography}
\newacronym{lvfo}{LVFOs}{Low-Voltage Fast Oscillations}
\newacronym{stft}{STFT}{Short Time Fourier Transform}
\newacronym{dft}{DFT}{Discrete Fourier Transform}
\newacronym{wt}{WT}{Wavelet Transform}
\newacronym{ei}{EI}{Epileptogenic Index}
\newacronym{hmm}{HMM}{Hidden Markov Model}
\newacronym{meg}{MEG}{Magneto-Electroencephalography}
\newacronym{kl}{KL}{Kullback–Leibler}
\newacronym{em}{EM}{Expectation-Maximization}
\newacronym{bw}{BW}{Baum–Welch}
\newacronym{auroc}{AUROC}{Area Under the Receiver Operating Characteristic Curve}
\newacronym{auprc}{AUPRC}{Area Under the Precision-Recall Curve}
\newacronym{cpparafac}{CP/PARAFAC}{Canonical Polyadic / Parallel Factors}
\newacronym{als}{ALS}{Alternate Least Square}

\usepackage{enumitem}
\setlist{nosep, leftmargin=14pt}

\usepackage{mwe} 

\title{Reading Across Space, Time and Frequencies: Tensor Decomposition of SEEG Signals to Support Epileptogenic Zone Definition}

\name{
Federico Mason$^{\dagger}$, Matteo Neri$^{\ddagger}$,
Maria Fratello$^{\dagger}$, Julia Makhalova$^{*,\dagger}$}

\address{\textit{Fabrice Bartolomei$^{*,\dagger}$,
Giovanni Petri$^{\S}$, Christian G. Bénar$^{\dagger}$}\\
$^{\dagger}$Institut de Neurosciences des Systèmes, Aix-Marseille University, Marseille, France\\
$^{\ddagger}$Institut de Neurosciences de la Timone, Aix-Marseille University, Marseille, France\\
$^{*}$Service de Neurophysiologie Clinique, Hôpital de la Timone, Marseille, France\\
$^{\S}$Network Science Institute, Northeastern University, London, United Kingdom}

\begin{document}
%
\maketitle

\begin{abstract}
Stereo-electroencephalography (SEEG) is an invasive technique used in the presurgical evaluation of drug-resistant epilepsy to support the localization of the Epileptogenic Zone (EZ).
SEEG data are particularly complex to analyze, as they include hundreds of components, named \emph{channels}, each associated with a different brain site, with interactions varying in both \emph{time} and \emph{frequency}.
In this study, we propose a new framework for SEEG analysis, based on the Canonical Polyadic (CP) decomposition, to extract the latent modes underlying seizure generation.
We model SEEG signals as a third-order tensor (channel × time × frequency), and factorize the tensor into rank-$1$ factors.
We then exploit the channel-specific factor as an unsupervised feature set for identifying the EZ.
The framework is validated on SEEG-recorded seizures from $35$ patients with drug-resistant epilepsy.
The results show that the joint combination of multiple frequency bands significantly enhances classification performance.
\end{abstract}


\section{Introduction}

\begin{tikzpicture}[remember picture, overlay]
\node[draw, minimum width=4in, font=\small, text=black] at ([yshift=-1cm]current page.north)  {This paper has been submitted to the IEEE International Symposium on Biomedical Imaging (ISBI) 2027.};
\end{tikzpicture}

The goal of epilepsy surgery is to resect the \gls{ez}, defined as the cortical area indispensable for the generation of seizures \cite{rosenow2001}.
When non-invasive evaluation is inconclusive, \gls{ez} location can be inferred via \gls{seeg}, an invasive procedure to record electrical activity from cortical and subcortical structures through intracranial electrodes with multiple contacts.
The result is a multi-dimensional signal, in which each channel reflects the activity of a brain site.
In clinical practice, \gls{seeg} recordings are reviewed by highly specialized neurophysiologists, with particular attention to the transition from the \emph{inter-ictal} (between seizures) to the \emph{ictal} period (during seizures) \cite{bartolomei2018interpretation}.
This analysis, based on visual inspection, allows one to identify the \gls{soz}, the region where seizures originate, and a key proxy for the \gls{ez}.

In recent years, many techniques have been developed to support \gls{seeg} interpretation, analyzing the spectral energy of single channels or functional connectivity between them.
The second option is particularly relevant given that seizures can originate from the interaction of distributed brain structures \cite{bartolomei2017defining}.
In particular, \gls{seeg} signals can be modeled as time-varying graphs, where nodes represent channels and edges represent connectivity relationships.
A variety of techniques have been proposed to estimate \gls{seeg} connectivity, with results that change depending on the frequency band analyzed. 
The \gls{soz} channels tend to desynchronize within the first seconds of seizure, considering slow oscillations \cite{mason2026desynchronization}, and increase high-frequency connectivity as ictal discharges propagate through the brain \cite{wendling2003epileptic}.
Hence, integrating connectivity information across different frequency bands is fundamental to provide a comprehensive description of \gls{soz}.

To cope with the inherent complexity of \gls{seeg} data, we can apply dimensionality reduction techniques to discover latent patterns behind brain activity.
In this context, \cite{frusque2020semi} proposed a joint matrix factorization to identify connectivity subgraphs shared across different seizures.
More recently, \cite{nieto2024epileptic} applied Dynamic Mode Decomposition (DMD) to separate frequency-specific modes involved in seizure generation.
However, these approaches do not integrate information from different oscillations: in \cite{frusque2020semi} connectivity is computed in a single frequency band, whereas in \cite{nieto2024epileptic} frequency is only used to sort the decomposition modes.
To jointly capture the effect of different frequency components, \gls{seeg} signals can be represented as tensors whose modes are nodes (or links), time, and frequency \cite{cong2015tensor, sidiropoulos2017tensor}.
This strategy was explored in \cite{li2018scalable}, which builds wavelet-power tensors of inter-ictal \gls{seeg} activity, and in \cite{chantal2021dynamic}, which considers connectivity tensors during seizures.
Nevertheless, these studies were validated in very small cohorts, and none of them assessed whether tensor modeling can support the identification of the \gls{ez}.

Inspired by the previous works in the literature, we develop a framework where the features of \gls{seeg} signals are modeled as a channel × time × frequency tensor.
We then apply the \gls{cpparafac} decomposition to factorize the tensor into a reduced number of components.
The factors associated with the spatial dimension are then exploited as an unsupervised feature set to identify the \gls{soz} channels.
We validate the framework on ictal data from $35$ patients with drug-resistant epilepsy, who underwent \gls{seeg} monitoring at La Timone Hospital (Marseille, France).
To the best of our knowledge, this is the first study applying tensor decomposition to ictal \gls{seeg} data from a large patient cohort.
Our results show that the integration of multiple frequencies into a single tensor model improves classification performance, suggesting that our framework is a promising tool for analysis of \gls{seeg} signals.

\section{System Model}
\subsection{Signal Model}
\label{sec:signal}
We model an \gls{seeg} recording as a multivariate discrete-time signal $\mathbf{s}[n] = \left\{s_c[n]\right\}_{c \in \mathcal{C}}$, $n = 0, \dots, N-1$, where $s_c[n] \in \mathbb{R}$ is the value of channel $c \in \mathcal{C}$ at sample $n$. Given the sampling frequency $f_s$ and the recording duration $T$, the total number of samples is $N = f_s \cdot T$.
We consider a set $\mathcal{Q}$ of frequency bands, with each band $q \in \mathcal{Q}$ defined by the interval $[f^q_{\text{low}}, f^q_{\text{high}})$ in Hz, with $0 \le f^q_{\text{low}} < f^q_{\text{high}} \le f_s/2$.
Hence, we denote by $s_c^q[n]$ the signal of channel $c$ filtered in band $q$.
The signal is segmented into overlapping windows of duration $\tau$, with a time shift $\Delta\tau < \tau$ between consecutive windows.
We choose $\tau$ and $\Delta\tau$ as multiples of the sampling period $1/f_s$, so that each window contains an integer number of samples $K = f_s \cdot \tau$, and consecutive windows are spaced by $M = f_s \cdot \Delta\tau$ samples.
The total number of windows is therefore $W = \lfloor (T - \tau)/\Delta\tau \rfloor + 1$, and window $w \in {0, 1, \dots, W-1}$ contains the samples from $w \cdot M$ to $w \cdot M + K - 1$.
Finally, we denote by $x_c^q[w,k]$ the $k$-th sample of channel $c$ within the $w$-th window for band $q$, i.e., $x_c^q[w,k] = s_c^q[w \cdot M + k]$, with $k \in {0, 1, \dots, K-1}$.

From the windowed signal, we now extract a set of features that characterize it through both local and connectivity-based measures.
First, for each channel $c$, window $k$ and frequency band $b$, we compute the energy
\begin{equation}
E_c^q[w] = \sum_{k=0}^{K-1} \left| x_c^q[w,k] \right|^2 .
\end{equation}
Then, we compute the instantaneous phase of $x_c^q[w,k]$ as the argument of its analytic signal:
\begin{equation}
\theta_c^q[w,k] = \arg\left( x_c^q[w,k] + j \cdot \mathcal{H}\left\{x_c^q[w,\cdot]\right\}[k] \right),
\end{equation}
where $\mathcal{H}\left\{ \cdot \right\}$ denotes the Hilbert transform \cite{oppenheim1989discrete}.
We partition the phase domain $[-\pi, \pi)$ into $L = \lceil \log_2 K \rceil + 1$ uniform bins $\vartheta_0,$ $\dots,$ $\vartheta_{L-1}$, according to Sturges' rule \cite{sturges1926choice}, and compute the entropy of channel $c$, window $w$, and band $q$, as
\begin{equation}
H_c^q[w] = - \sum_{\ell=0}^{L-1} p_c^q[\ell; w] , \log_2 p_c^q[\ell; w],
\end{equation}
where $p_c^q[\ell; w] = \frac{1}{M} \left| { k : \theta_c^q[w, k] \in \vartheta_\ell } \right|$ is the empirical probability that the phase samples fall into bin $\vartheta_\ell$.
Finally, for window $w$ and a frequency band $q$, we measure the connectivity between channels $c$ and $h$ through the \gls{mi} between the phase distributions:
\begin{equation}
I_{c,h}^q[w] = H_c^q[w] + H_h^q[w] - H_{c,h}^q[w],
\end{equation}
where $H_c^q[w]$ and $H_h^q[w]$ are defined as above, and $H_{c,h}^q[w]$ is the joint entropy of the two distributions \cite{shannon1948mathematical}.

\subsection{Tensor Decomposition}
\label{sec:tensor}

We consider three tensor models of the \gls{seeg} signal, depending on the feature used to characterize it:
\begin{itemize}
\item an energy tensor $\boldsymbol{\mathcal{X}}_E \in \mathbb{R}^{C \times W \times Q}$, whose entries are the signal energies of each channel, window, and frequency band, i.e., $\boldsymbol{\mathcal{X}}_E[c,w,q] = E_c^q[w]$;
\item an entropy tensor $\boldsymbol{\mathcal{X}}_H \in \mathbb{R}^{C \times W \times Q}$, whose entries are the phase entropies, i.e., $\boldsymbol{\mathcal{X}}_H[c,w,q] = H_c^q[w]$;
\item a \gls{mi} tensor $\boldsymbol{\mathcal{X}}_I \in \mathbb{R}^{C^2 \times W \times Q}$, whose entries are the functional connectivity of each channel pair, window and frequency band, i.e., $\boldsymbol{\mathcal{X}}_I[p,w,q] = I_{c,h}^q[w]$, where $c = \lfloor p/C \rfloor$ and $h = p \bmod C$.
\end{itemize}
To ease the reading, in the following we refer to $\boldsymbol{\mathcal{X}}_E$ and $\boldsymbol{\mathcal{X}}_H$, but the same formulation applies to $\boldsymbol{\mathcal{X}}_I$, with the first mode indexing channel pairs instead of channels.
Following the \gls{cpparafac} approach \cite{bro1997parafac}, we approximate the tensor $\boldsymbol{\mathcal{X}} \in \mathbb{R}^{C \times W \times Q}$ as a sum of $R$ non-negative rank-$1$ tensors:
\begin{equation}
\label{eq:decomp}
\hat{\boldsymbol{\mathcal{X}}} = \sum_{r=0}^{R-1} \lambda_r \cdot \mathbf{c}_r \circ \mathbf{w}_r \circ \mathbf{q}_r,
\end{equation}
where $\circ$ denotes the vector outer product, $\lambda_r \in \mathbb{R}$, $\mathbf{c}_r \in \mathbb{R}^{C}$, $\mathbf{w}_r \in \mathbb{R}^{W}$ and $\mathbf{q}_r \in \mathbb{R}^{Q}$.
This representation decomposes the \gls{seeg} signal into the superposition of $R$ latent brain components. The weight $\lambda_r$ quantifies the importance of component $r$, while the vectors $\mathbf{c}_r$, $\mathbf{w}_r$ and $\mathbf{q}_r$ are its spatial (channel), temporal (window) and spectral (frequency) signatures.

To solve \eqref{eq:decomp}, we employ the \gls{als} algorithm.
We first collect the factor vectors of the $R$ components into the factor matrices $\mathbf{C} \in \mathbb{R}^{C \times R}$, $\mathbf{W} \in \mathbb{R}^{W \times R}$ and $\mathbf{Q} \in \mathbb{R}^{Q \times R}$, whose $r$-th columns are $\mathbf{c}_r$, $\mathbf{w}_r$ and $\mathbf{q}_r$, respectively. The algorithm alternately updates each factor matrix, while the other two are fixed.
In particular, the update of $\mathbf{C}$ is
\begin{equation}
\label{eq:als_0}
\mathbf{C} \leftarrow \mathbf{X}_{(0)} \left( \mathbf{W} \odot \mathbf{Q} \right) \left( \left( \mathbf{W}^T \mathbf{W} \right) \ast \left( \mathbf{Q}^T \mathbf{Q} \right) \right)^{-1},
\end{equation}
where $\odot$ is the Khatri-Rao product, $\ast$ is the Hadamard (element-wise) product, and $\mathbf{X}_{(0)} \in \mathbb{R}^{C \times W\cdot Q}$ is the mode-$0$ unfolding of the tensor, defined as $\mathbf{X}_{(0)}[c, w \cdot Q + q] = \boldsymbol{\mathcal{X}}[c,w,q]$.
The matrices $\mathbf{W}$ and $\mathbf{Q}$ are updated analogously, using the corresponding unfoldings of $\boldsymbol{\mathcal{X}}$.
Hence, each column of the factor matrices is normalized to unit $\ell_2$ norm, and the norms are absorbed into the weights:
\begin{equation}
\lambda_r = | \mathbf{C}[:,r] |_2 \cdot | \mathbf{W}[:,r] |_2 \cdot | \mathbf{Q}[:,r] |_2 .
\label{eq:als_4}
\end{equation}
The updates \eqref{eq:als_0}-\eqref{eq:als_4} are repeated until convergence, i.e., until the relative change of the reconstruction error $| \boldsymbol{\mathcal{X}} - \hat{\boldsymbol{\mathcal{X}}} |_F / | \boldsymbol{\mathcal{X}} |_F$ falls below a predefined tolerance.

\subsection{Unsupervised Feature Extraction}
\label{sec:classif}
We assume that the factors $\{\mathbf{c}_r\}_{r=0}^{R-1}$ capture the spatial organization of the \gls{seeg} signal, and can therefore be used to characterize epileptogenic channels.
In the energy-based tensor, each channel $c \in \mathcal{C}$ is associated with the vector $\mathbf{d}_E(c) = \mathbf{C}[c,:] \in \mathbb{R}^{R}$, describing the participation of $c$ in the $R$ latent components of the signal.
Hence, we can train a classifier that takes $\mathbf{d}_E(c)$ as input and outputs the class (\gls{ez} vs non-\gls{ez}) of channel $c$.
In the case of the entropy-based tensor, we follow the same approach, using $\mathbf{d}_H(c) = \mathbf{C}[c,:] \in \mathbb{R}^{R}$ as the classification input.
We note that the dimensions of $\mathbf{d}_E(c)$ and $\mathbf{d}_H(c)$ do not depend on the number of windows $W$ or frequency bands $B$, but only on the rank $R$. 

\begin{figure}[t!]
\centering
\includegraphics[width=.9\linewidth]{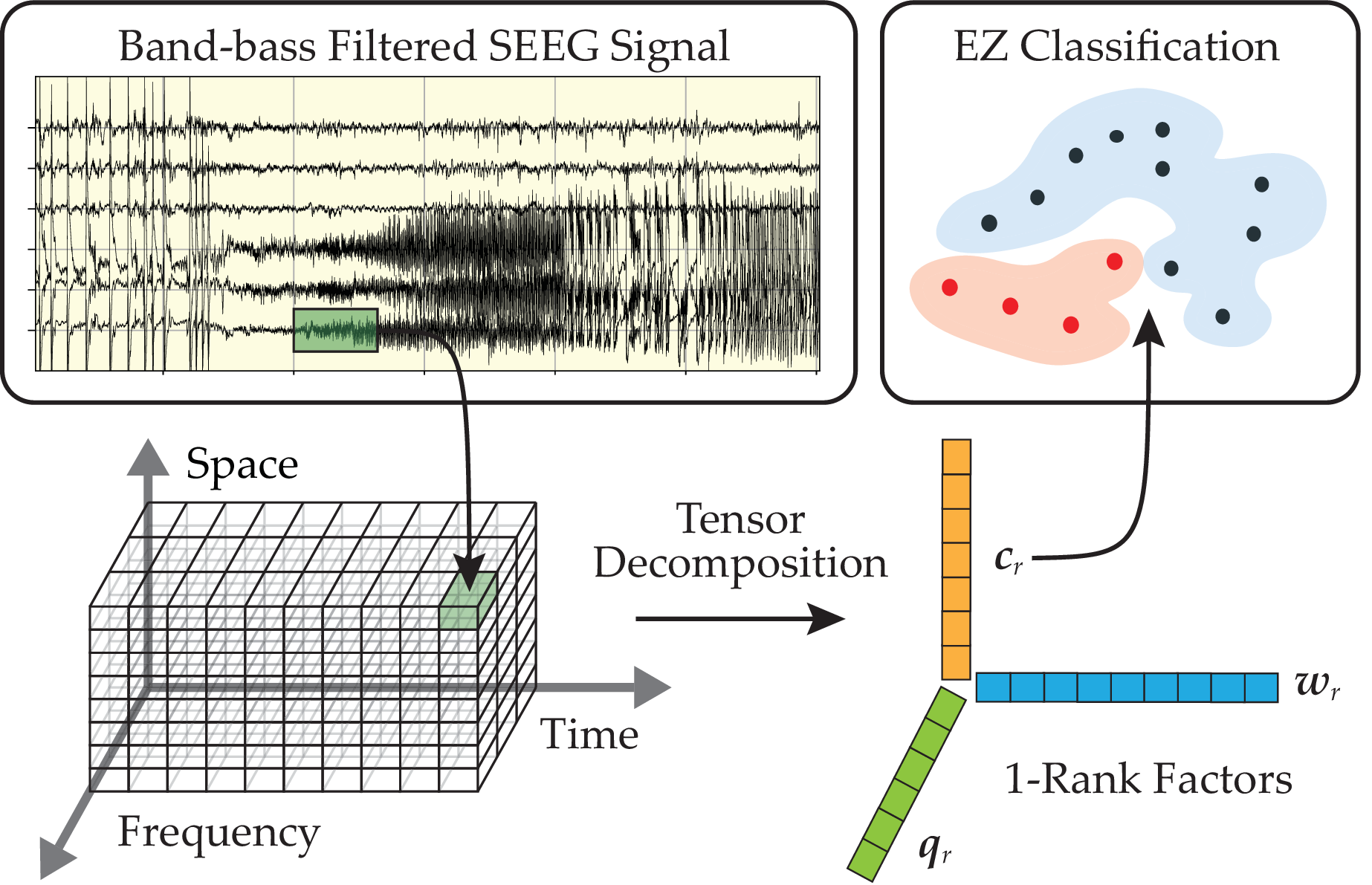}
\caption{Methodological pipeline\vspace{-0.4cm}.}
\label{fig:tensor}
\end{figure}

With the \gls{mi}-based tensor, the rows of $\mathbf{C}$ are associated with channel pairs, so there is no one-to-one correspondence between channels and the decomposition output.
To obtain channel-level features, we aggregate the links incident to each channel.
Let $\mathcal{L}(c) = \{ p: p = c \cdot C + h \vee p = h \cdot C + c, h = 0, \cdots, C - 1  \}$ be the set of links involving channel $c$. For each component $r$ and each percentile threshold $\eta \in \mathcal{H}$, we define the thresholded weighted degree of $c$ as
\begin{equation}
d_{\eta}(c,r) = \sum_{p \in \mathcal{L}(c)} \mathbf{C}[p, r] \cdot \mathbb{I}\left( \mathbf{C}[p, r] > P_{\eta}\left(\mathbf{C}[:, r]\right) \right),
\end{equation}
where $\mathbb{I}(\cdot)$ is the indicator function and $P_{\eta}(\mathbf{C}[:, r])$ is the $\eta$-th percentile of the links associated with the $r$-th component. 
The unsupervised feature vector is finally obtained as
\begin{equation}
\mathbf{d}_I(c) = \operatorname*{concat}_{\eta \in \mathcal{H}} \left( d_{\eta}(c,0), \dots, d_{\eta}(c,R-1) \right) \in \mathbb{R}^{R \cdot |\mathcal{H}|}.
\end{equation}
A schematic representation of the proposed methodological pipeline is shown in Fig.~\ref{fig:tensor}.

\section{Experimental Results}

\subsection{Dataset and Settings}
We retrospectively analyzed \gls{seeg} data of $35$ patients from the EPINOV database of La Timone Hospital, (Marseille, France).
The patients were selected according to the following inclusion criterion: at least one seizure starting with low-voltage fast activity.
The signals were sampled at $f_s=500$ Hz and then converted to a bipolar montage.
For each patient, we selected an epoch of $T=50\,\mathrm{s}$ spanning $35\,\mathrm{s}$ s before and $15\,\mathrm{s}$ after the electrographic seizure onset, i.e., the start of fast oscillations.
We segment each epoch into windows lasting $\tau=\,\mathrm{s}$, with a shift of $\Delta\tau=0.25\,\mathrm{s}$.

Following the model in Sec. \ref{sec:signal}, we computed the energy $E_c^q[w]$, entropy $H_c^q[w]$, and connectivity $I_{c,h}^q[w]$ for each channel $c$, link $(c,h)$, window $w$, and frequency band $q \in \mathcal{Q}=\{\delta, \theta, \alpha, \beta, \gamma_{\text{low}}, \gamma_{\text{high}}\}$, where $\delta = [2, 4),$ $\theta = [4, 8),$ $\alpha = [8, 14),$ $\beta = [14, 30),$ $\gamma_{\text{low}} = [30, 70),$ $\gamma_{\text{high}} = [70, 150)\, \text{Hz}$.
To capture ictal-related changes for each feature, we applied a $Z$-score normalization using the first $30\,\mathrm{s}$ (i.e., the first $120$ windows) of each recording to estimate the mean ($\mathbb{E}$) and standard deviation ($\operatorname{STD}$) of each feature.
For energy, this yields
\begin{equation}
\hat{E}_c^q[w] = \frac{E_c^q[w] - \mathbb{E}_{w<120}[E_c^q[\cdot]]}{\operatorname{STD}_{w<120}[E_c^q[\cdot]]},
\end{equation}
with analogous expressions for entropy and \gls{mi}.

Following the model in Sec. \ref{sec:tensor}, we then constructed the tensors $\boldsymbol{\mathcal{X}}_E$, $\boldsymbol{\mathcal{X}}_H$, and $\boldsymbol{\mathcal{X}}_I$ from the remaining $20\,\mathrm{s}$ of each epoch, with the seizure starting after $5$ s.
The tensor dimensions are determined by the number $C$ of channels, which is patient-specific, the number $W$ of windows, and the number $Q$ of frequency bands.
In our experiments, we vary the portion of the ictal signal analyzed and the composition of the frequency band set $\mathcal{Q}$, obtaining different results for $W \in \left[1, W_{\text{max}}\right]$ and $Q\in\{1,\ldots,6\}$, where $W_{\text{max}}=77$ denotes the last window of the epoch.
Finally, we set the rank of the decomposition to $R=6$, and the percentiles used to compute the weighted degrees to $\mathcal{H}=\{75, 80, 85, 90, 95\}$.

We used the unsupervised feature vectors $\mathbf{d}_E(c)$, $\mathbf{d}_H(c)$, and $\mathbf{d}_I(c)$ as inputs to a logistic regression classifier to identify channels labeled as part of the \gls{soz}.
Ground truth was defined by the clinical annotations provided by neurophysiologists following visual inspection of the \gls{seeg} recordings.
Classification performance was assessed using the \gls{auroc} and the \gls{auprc}.
Our classification task presents strongly imbalanced classes and, thus, the \gls{auprc} is a preferable metric for system evaluation\cite{fawcett2006introduction}. 
Since the goal of this study is to assess the intrinsic discriminative power of the unsupervised features, performance was evaluated in a within-patient setting, while cross-patient generalization is left for future work.

\begin{figure}[t]
    \centering    \includegraphics[width=.9\linewidth]{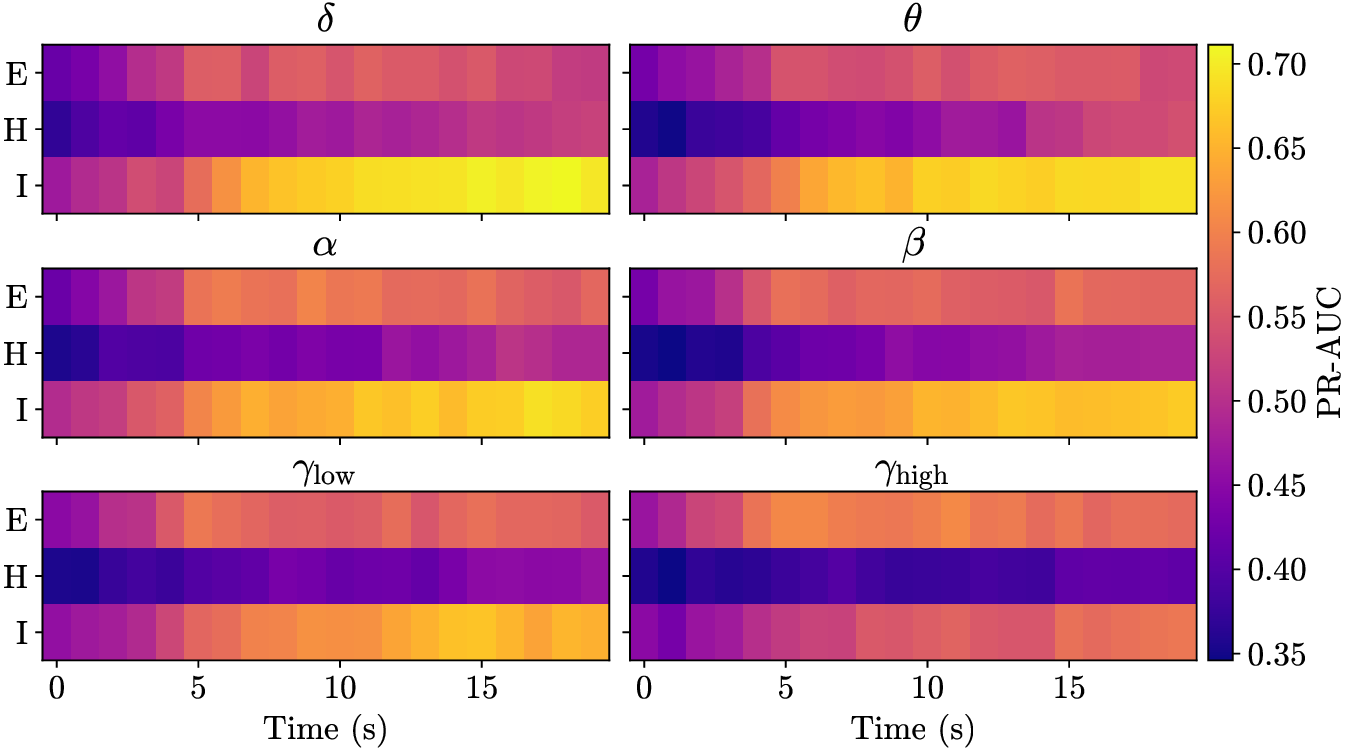}
    \caption{Average PR-AUC in the patient cohort as a function of time, using a single frequency band in the tensor model\vspace{-0.2cm}.}
    \label{fig:pr_auc_single_band}
\end{figure}

\begin{figure}[t]
\centering
    \begin{subfigure}{.9\linewidth}
    \centering
    \includegraphics[width=\linewidth]{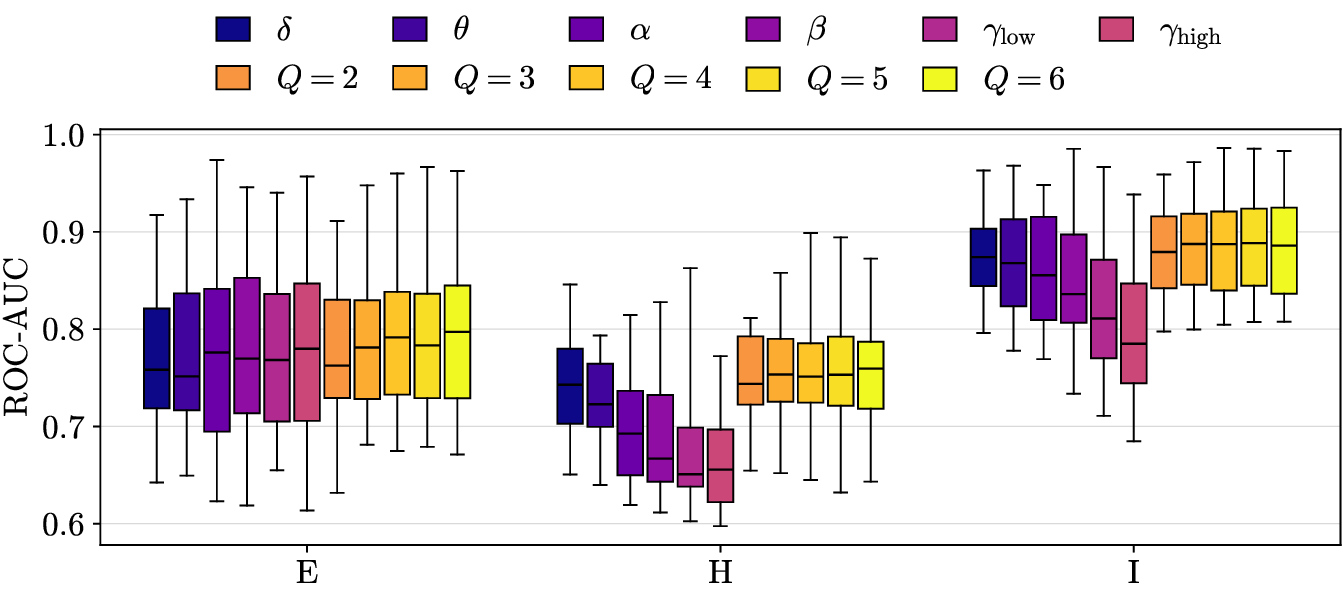}
    \caption{Time-averaged ROC-AUC.}
    \label{fig:roc_auc_box}
    \end{subfigure}\\
    \begin{subfigure}{.9\linewidth}
    \centering
    \includegraphics[width=\linewidth]{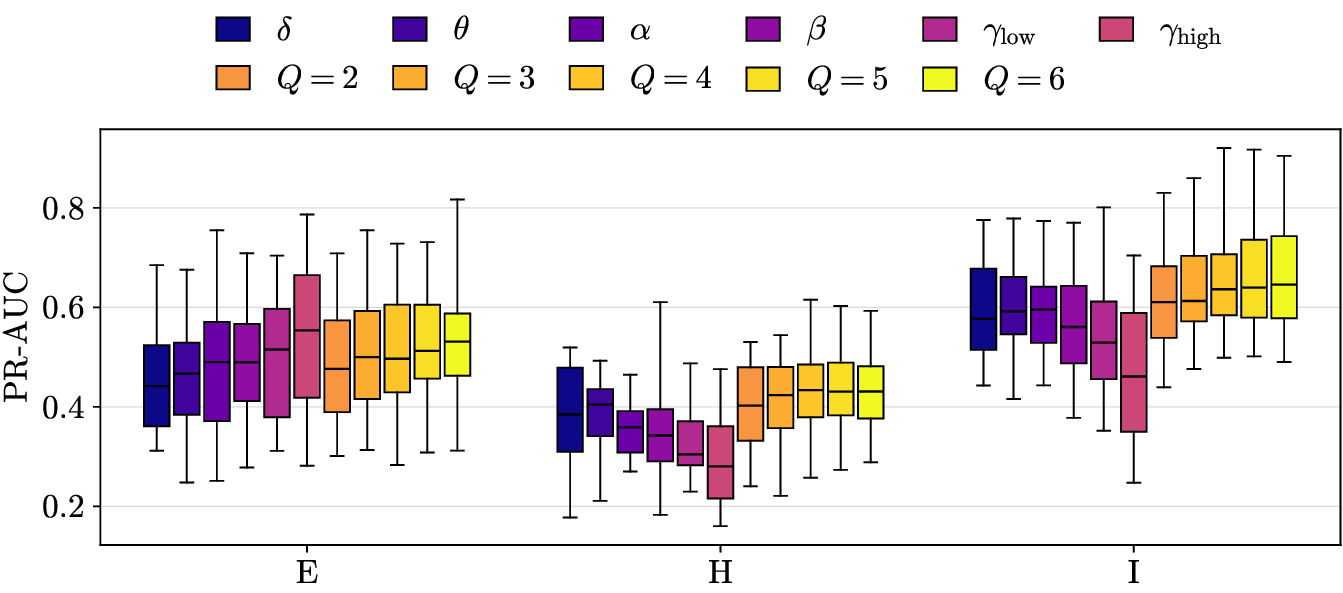}
    \caption{Time-averaged PR-AUC.}
    \label{fig:pr_auc_box}
    \end{subfigure}    
    \caption{Time-averaged performance distribution in the patient cohort for different frequency sets $\mathcal{Q}$\vspace{-0.2cm}.}
    \label{fig:performance_box}
\end{figure}

\subsection{Numerical Results}

We first consider the edge scenario where the feature set $\mathcal{Q}$ includes a single frequency band, so that the tensor has shape $C \times W \times 1$ for energy and entropy, and $C^2 \times W \times 1$ for \gls{mi}.
Fig.~\ref{fig:pr_auc_single_band} reports the \gls{auprc}, averaged across the patient cohort, as a function of the feature used to build the tensor and of the extension of the signal analyzed.
We note that performance exhibits different trends depending on the feature used to build the tensor.
The energy-based model achieves its best results during the first seconds of the seizure (which starts after $5$ s), particularly when using the $\alpha$ and the $\gamma_{\text{high}}$ bands.
Instead, the entropy-based model performs best in the low portion of the frequency spectrum and at later stages of the seizure.
Finally, the \gls{mi}-based model yields the highest \gls{auprc} overall, with maxima in the $\delta$ and $\theta$ bands and approximately $5$ s after seizure onset.

Fig.~\ref{fig:performance_box} reports the distribution across patients of the time-averaged \gls{auroc} and \gls{auprc} as a function of the frequency bands included in the tensor.
Each distribution is shown as a box plot, where the central line is the median, the edges are the first and third quartiles, and the whiskers are the $5$-th and $95$-th percentiles.
We denote by $Q=2$ the configuration with the $\delta$ and $\theta$ bands and, for increasing $Q$, the configurations obtained by adding one band at a time.
For the energy-based tensor, the best performance is obtained with $\gamma_\text{high}$, while including additional bands does not improve either \gls{auroc} or \gls{auprc}.
The entropy-based tensor shows the opposite trend: the best performance is obtained with the low-frequency bands, and it decreases monotonically as higher portions of the spectrum are included.

For the \gls{mi}-based tensor, the performance trends are similar to those of entropy, even though with higher \gls{auroc} and \gls{auprc} values in all frequency configurations.
In this case, combining multiple bands yields more robust unsupervised features: all the multi-band configurations improve the \gls{auroc} and \gls{auprc} in both the upper tails with respect to the single-band models.
In particular, the configuration integrating all frequencies ($Q=6$) reaches mean \gls{auroc} and \gls{auprc} values of $0.887$ and $ 0.655$, respectively.
Performing a one-sided Wilcoxon signed-rank test \cite{wilcoxon1992individual}, with the alternative hypothesis that $Q=6$ achieves a higher \gls{auprc} than single-band tensors, we obtain significant differences in all comparisons.
The best single-band configuration ($\theta$) is outperformed with a p-value $p = 0.001$, while all remaining configurations are outperformed with $p < 0.001$.

\section{Conclusions}

In this work, we proposed a new framework to model \gls{seeg} signals across space (channels), time, and frequency.
We built three \gls{seeg} tensor models that characterize the signal through its energy, entropy, and functional connectivity.
We applied the \gls{cpparafac} approach to decompose the tensor into rank-$1$ factors, and exploited the spatial factor as an unsupervised feature set for \gls{ez} definition. 
The framework was evaluated on a dataset of $35$ patients who underwent \gls{seeg} monitoring at La Timone Hospital (Marseille, France).

Our results show that the performance trends across frequency bands depend on the feature used to build the tensor, and that integrating multiple frequency bands in the same model improves the identification of the \gls{ez} in terms of \gls{auroc} and \gls{auprc}.
Future work will address the validation of the framework on an independent patient cohort, using cross-validation strategies to assess the generalizability of the results beyond the within-patient settings.

\section{Compliance with Ethical Standards}
This retrospective study exclusively used pseudonymised data acquired as part of EPINOV (NCT03643016) prospective study. All subjects gave their informed written consent to participate in the study according to the Declaration of Helsinki. In accordance with French regulations on retrospective research using health data, all patients received written information about the reuse of their data for research and did not object (non-opposition procedure). No potentially identifying information is reported.

\section{Acknowledgments}
Part of this work was supported by the European Union under the Marie Skłodowska-Curie Actions (MSCA) postdoctoral fellowships program of Horizon Europe, Grant Agreement No. 101270737 (HYPHEN).
Part of this work received funding from the French National Research Agency (ANR) under the second “Investissements d’Avenir” program (ANR-17-RHUS-0004, EPINOV).

\bibliographystyle{IEEEtran}
\bibliography{refs.bib}

\end{document}